# ПРИМЕНЕНИЕ НЕЛИНЕЙНОГО ОПЕРАТОРА ДЛЯ ИДЕНТИФИКАЦИИ НЕИЗВЕСТНОГО ПАРАМЕТРА ДЛЯ СКАЛЯРНОГО РЕГРЕССИОННОГО УРАВНЕНИЯ С ПОМЕХОЙ В КАНАЛЕ ИЗМЕРЕНИЯ

В.С.Воробьев, А.А.Бобцов, Николаев Н.А., Пыркин А.А.
Университет ИТМО, Санкт-Петербург, 197101, Российская Федерация
Адрес для переписки: v.s.vorobyev@yandex.ru

**Аннотация**
**Предмет исследования.** В статье исследуется алгоритм идентификации неизвестного постоянного параметра для скалярной регрессионной модели с применением нелинейного оператора, позволяющего получить новое регрессионное уравнение (с расширенным числом неизвестных параметров), для которого влияние помех в измерении или возмущающего воздействия будет минимальным. **Целью работы** является разработка нового метода идентификации неизвестного постоянного параметра для классической линейной регрессионной модели при наличии помех измерения или возмущающих воздействий. Поставленная задача решается при условии, что модуль амплитуды помехи или возмущающего воздействия меньше единицы и, более того, не превосходит полезную составляющую измеряемого сигнала. **Методы.** Предложен метод оценивания параметра линейной скалярной регрессионной модели на базе применения нелинейного оператора (в рамках данной статьи была выбрана экспоненциальная функция/оператор), позволяющего расширить регрессионное уравнение до нескольких неизвестных параметров, но при этом нивелировать влияние помех или возмущений. Метод основан на разложении экспоненциальной функции в ряд Тейлора с отсеканием части членов с существенно малой амплитудой, а также последующего применения метода динамического расширения и смешивания регрессора, обеспечивающего возвращение к скалярной регрессионной модели, что в свою очередь, обеспечивает равномерную сходимость и высокое быстродействие при использовании градиентных методов идентификации. **Основные результаты.** Предлагаемый метод позволяет при наличии помех измерения повысить точность оценки параметра регрессионной модели по сравнению с классическим методом. Более того, предлагаемый подход дает методику расширения регрессионного уравнения, обеспечивающего уменьшение влияния помех. Иными словами, чем больше становится параметров в новой регрессионной модели (полученной при использовании нелинейного оператора для исходной регрессии), тем меньше значение шума или возмущения. **Практическая значимость.** Предлагаемый в статье метод является новым рабочим инструментом в задачах идентификации неизвестных постоянных параметров. Областью применения данного метода является идентификация параметров математических моделей систем, сводимых к виду линейных регрессионных уравнений, содержащих помехи измерений или возмущающие воздействия. Данный подход может использоваться для широкого класса задач управления техническими объектами, для которых актуальна задача идентификации неизвестных постоянных параметров.
**Ключевые слова:** идентификация параметров, линейная регрессия, нелинейный оператор, помехи в измерениях.

## Введение

Статья посвящена классической задаче идентификации параметров для скалярной линейной регрессионной модели, то есть статического уравнения левая часть, которого известна, а правая содержит сумму из *n* неизвестных постоянных параметров, умноженных на *n* известных функций – регрессоров. Существует множество подходов к оценке параметров линейной регрессионной модели (большинство из них можно найти в [1]). Если линейное регрессионное уравнение не содержит помех в измерениях или возмущений в своей правой части, то при условии незатухающего возбуждения на регрессоры (см., например, [2] и [3]) в случае использования метода градиентного спуска параметры будут найдены асимптотически точно. Однако при наличии шумов измерений или возмущающих воздействий оценивание параметров будет осуществляться с ошибкой.

В целом методы идентификации параметров можно разделить на две большие группы: с постобработкой данных и реального времени. Методы, использующие постобработку данных, допускают применение более сложных алгоритмов и предусматривают знание статистических величин, полученных в результате многочисленных наблюдений за некоторым процессом (см. [4], [5]). Например, наблюдатель, построенный на основе фильтра Калмана запрашивает знания дисперсии возмущения [6]. Методы идентификации в реальном времени не требуют набора экспериментальных/статических данных, поэтому широко применяются в задачах оценивания параметров (см., например, [7], [8] и [9]).

Как уже указывалось ранее, одним из классических и широко распространенным подходов в идентификации является метод градиентного спуска, требующий для асимптотической сходимости по настраиваемым параметрам выполнения условия незатухающего возбуждения [10]. В работе [11] отмечено, что настройка градиентного метода идентификации обычно заключается в подборе коэффициента усиления и требует множества попыток. При этом, как показано в [12], увеличение коэффициента усиления в методе градиентного спуска не всегда приводит к повышению скорости переходного процесса, но при этом ведет к росту пульсации и обострению выбросов. Однако при наличии помех или возмущений изменение коэффициента усиления на точность сходимости оценок параметров к истинным значениям не оказывает существенного влияния. Для скалярных регрессионных моделей (то есть для уравнений с одним неизвестным параметром) для высокочастотных помех уменьшение коэффициента усиления может привести к улучшению точности, но при этом увеличивается время сходимости. При этом увеличение коэффициента усиления приводит к увеличению быстродействия, но ухудшению точности.

В этой работе предлагается новый подход, позволяющий увеличить точность оценивания параметра скалярной регрессионной модели при наличии возмущения, основанный на применении нелинейного экспоненциального оператора и перепараметризации существующей регрессионной модели. Предлагаемый подход позволяет исключить рост влияния возмущения при увеличении коэффициента усиления при идентификации параметра с использованием метода градиентного спуска.

## Постановка задачи

Рассмотрим скалярную регрессионную модель вида:
$$\bar{y}(t) = \bar{\phi}(t)\theta + \bar{\delta}(t), \qquad (1)$$
где $\bar{y}(t)$ и $\bar{\phi}(t)$ – измеряемые сигналы, $\theta$ – неизвестный постоянный параметр, $\bar{\delta}(t)$ – некоторая неизвестная ограниченная функция времени или помеха измерений.

Ставится задача синтеза алгоритма оценки $\hat{\theta}(t)$ неизвестного параметра $\theta$, чтобы:

1) при $\bar{\delta}(t) = 0$ выполнялось
$$\lim_{t \to \infty} \left\| \theta - \hat{\theta}(t) \right\| = 0, \qquad (2)$$
где $\|\cdot\|$ – евклидова норма;

2) при ненулевой функции $\bar{\delta}(t)$ обеспечивалось неравенство
$$\left| \theta - \hat{\theta}(t) \right| \leq \varepsilon_0, \qquad (3)$$
где $\varepsilon_0 > 0$ – в обще случае, некоторое малое число.

Сформулированная задача будет решаться при следующих допущениях.

**Допущение 1.** Помеха измерений $\bar{\delta}(t)$ строго меньше модуля полезного сигнала $\bar{\phi}(t)\theta$.

**Допущение 2.** Функция $\bar{\delta}(t)$ такая, что выполняется $\left|\bar{\delta}(t)\right| \leq 1$.

## Основной результат

К регрессионной модели (1) применим линейный оператор/фильтр
$$\frac{k}{p+k},$$
где $p := \frac{d}{dt}$ – оператор дифференцирования, $k > 0$ – коэффициент усиления фильтра.

В результате уравнение (1) примет вид:
$$y(t) = \phi(t)\theta + \delta(t), \qquad (4)$$
где $y(t) = \frac{k}{k+p}\bar{y}(t)$, $\phi(t) = \frac{k}{k+p}\bar{\phi}(t)$, $\delta(t) = \frac{k}{k+p}\bar{\delta}(t)$.

Для нахождения оценки $\hat{\theta}_{new}(t)$ неизвестного постоянного параметра $\theta$ применим к уравнению (4) нелинейный оператор вида
$$x(t) := e^{y(t)}. \qquad (5)$$

**Замечание 1.** Следует отметить, что в этой статьей в качестве нелинейного оператора была использована экспоненциальная функция, хотя данный выбор, на взгляд авторов, не является единственным и для преобразования уравнения (4) может использоваться любой оператор, позволяющий нивелировать влияние сигнала $\bar{\delta}(t)$.

Подставляя (4) в уравнение (5), получаем
$$x = e^y = e^{\phi\theta+\delta} = e^{\phi\theta}e^{\delta} = \left[1+\delta+\frac{\delta^2}{2!}+\ldots\right]e^{\phi\theta} \approx \left[1+\delta+\frac{\delta^2}{2}\right]e^{\phi\theta} = \qquad (6)$$
$$= \left[1+(y-\phi\theta)+\frac{1}{2}(y-\phi\theta)^2\right]e^{\phi\theta} = \left[\left(1+y+\frac{1}{2}y^2\right)-(\phi+\phi y)\theta+\left(\frac{1}{2}\phi^2\right)\theta^2\right]e^{\phi\theta},$$
где член $e^{\delta}$ был представлен в виде разложения в ряд Тейлора в окрестности нуля, а далее все слагаемые разложения $e^{\delta}$ со степенью больше квадрата были опущены (полагая их существенно малыми, согласно Допущению 2).

Для компактности записи введем новые обозначения:
$$\begin{cases} \alpha := 1+y+\frac{1}{2}y^2; \\ \tau := -\phi-\phi y; \\ \rho := \frac{1}{2}\phi^2. \end{cases} \qquad (7)$$

С учетом (7) выражение (6) примет вид:
$$x = \left[\alpha+\tau\theta+\rho\theta^2\right]e^{\phi\theta}. \qquad (8)$$

Откуда легко видеть, что
$$e^{\phi\theta} = \frac{x}{\alpha+\tau\theta+\rho\theta^2}. \qquad (9)$$

Найдем производную для (8):
$$\dot{x} = \left[\dot{\alpha}+\dot{\tau}\theta+\dot{\rho}\theta^2\right]e^{\phi\theta} + \left[\alpha+\tau\theta+\rho\theta^2\right]\theta\dot{\phi}e^{\phi\theta}. \qquad (10)$$

Подставляя (7) и (8) в уравнение (9), получаем:
$$\dot{x}\left[\alpha+\tau\theta+\rho\theta^2\right] = \dot{\alpha}x+\dot{\tau}x\theta+\dot{\rho}x\theta^2+\alpha\dot{\phi}x\theta+\tau\dot{\phi}x\theta^2+\rho\dot{\phi}x\theta^3. \qquad (11)$$

Откуда после раскрытия скобок имеем:
$$\alpha\dot{x} - \dot{\alpha}x = \left[\dot{\tau}x - \tau\dot{x} + \alpha\dot{\phi}x\right]\theta + \left[\dot{\rho}x - \rho\dot{x} + \tau\dot{\phi}x\right]\theta^2 + \rho\dot{\phi}x\theta^3. \tag{12}$$

Введем новые обозначения:
$$\begin{cases} q = \alpha\dot{x} - \dot{\alpha}x; \\ \psi_1 = \dot{\tau}x - \tau\dot{x} + \alpha\dot{\phi}x; \\ \psi_2 = \dot{\rho}x - \rho\dot{x} + \tau\dot{\phi}x; \\ \psi_3 = \rho\dot{\phi}x. \end{cases} \tag{13}$$

Поскольку
$$\dot{x} = \left(e^y\right)' = \dot{y}e^y = \dot{y}x,$$

то с учетом обозначений (7) для уравнения (13) имеем:
$$q = \left[1 + y + \frac{1}{2}y^2\right]\dot{x} - \left[1 + y + \frac{1}{2}y^2\right]'x = \dot{y}x + \dot{y}yx + \frac{1}{2}y^2\dot{x} - \dot{y}x - \dot{y}yx = \frac{1}{2}y^2\dot{x}; \tag{14}$$

$$\psi_1 = \left[-\phi - \phi y\right]'x - \left[-\phi - \phi y\right]\dot{x} + \left[1 + y + \frac{1}{2}y^2\right]\dot{\phi}x =$$
$$= -\dot{\phi}x - \dot{\phi}yx - \phi\dot{y}x + \phi\dot{y}x + \phi y\dot{y}x + \dot{\phi}x + y\dot{\phi}x + \frac{1}{2}y^2\dot{\phi}x = \phi y\dot{y}x + \frac{1}{2}y^2\dot{\phi}x;$$

$$\psi_2 = \left[\frac{1}{2}\phi^2\right]'x - \left[\frac{1}{2}\phi^2\right]\dot{x} + \left[-\phi - \phi y\right]\dot{\phi}x = -\frac{1}{2}\phi^2\dot{y}x - \phi y\dot{\phi}x;$$

$$\psi_3 = \frac{1}{2}\phi^2\dot{\phi}x.$$

где символ «′» обозначает производную.

Таким образом после всех преобразований получаем новую регрессионную модель вида
$$q = \psi_1\theta + \psi_2\theta^2 + \psi_3\theta^3 = \begin{pmatrix}\psi_1 & \psi_2 & \psi_3\end{pmatrix}\begin{pmatrix}\theta \\ \theta^2 \\ \theta^3\end{pmatrix} = \Psi^T\Theta, \tag{15}$$

где $\Psi^T = \begin{pmatrix}\psi_1 & \psi_2 & \psi_3\end{pmatrix}$ – новый вектор регрессии, $\Theta^T = \begin{pmatrix}\theta & \theta^2 & \theta^3\end{pmatrix}$ – вектор, состоящий из степеней оцениваемого параметра, $q$ – измеряемый сигнал.

**Замечание 2.** Следует отметить, что в новой регрессионной модели (15) отсутствуют какие-либо неучтенные помехи или шумы, вызванные сигналом $\bar{\delta}(t)$. Как было показано ранее, все слагаемые разложения $e^\delta$ со степенью больше квадрата были приняты пренебрежимо малыми. Таким образом вместо регрессионных моделей (1) и (4), содержащих помеху, получено уравнение (15), включающее в себя три неизвестных параметра, являющихся нелинейной комбинацией $\theta$.

Для получения оценки $\hat{\theta}_{new}(t)$ неизвестного параметра $\theta$ применим процедуру ДРСР (динамического расширения и смешивания регрессора, Dynamic Regressor Extension and Mixing, DREM), впервые предложенную в [13] (см. также [14] и [15] для подробностей и расширений).

Согласно процедуре ДРСР:

1. К (15) применим линейные устойчивые фильтры $B_i(p) = \dfrac{\beta_i}{p+\beta_i}$, $i = \overline{1,2}$ для получения расширенной регрессии вида:
$$Q_e = \Psi_e \Theta, \qquad (16)$$
где $Q_e = \begin{pmatrix} q \\ q_{1\,filtered} \\ q_{2\,filtered} \end{pmatrix}$, $\Psi_e = \begin{pmatrix} \Psi^T \\ \Psi^T_{1\,filtered} \\ \Psi^T_{2\,filtered} \end{pmatrix}$.

2. Умножим (16) на присоединенную матрицу $\Psi_e$ и получим:
$$Z = \Delta \Theta, \qquad (17)$$
где $\Delta = \det(\Psi_e)$, $Z = adj\{\Psi_e\} Q_e$.

Тогда оценка $\hat{\theta}_{new}(t)$ может быть найдена методом градиентного спуска из уравнения (17) следующим образом:
$$\dot{\hat{\theta}}_{new} = \kappa \Delta \left( z_1 - \Delta \hat{\theta}_{new} \right), \qquad (18)$$
где $\kappa > 0$ – задаваемый коэффициент усиления, $z_1 = \begin{pmatrix} 1 & 0 & 0 \end{pmatrix} Z$ – первый элемент вектора $Z$.

Заметим, что применение нового подхода не исключает использования классического варианта решения задачи идентификации неизвестного параметра для модели (1). В самом деле оценка $\hat{\theta}_{gradient}(t)$ из скалярной регрессии (4) может быть найдена с помощью градиентного метода:
$$\dot{\hat{\theta}}_{gradient} = \gamma_{gradient} \phi \left( y - \phi \hat{\theta}_{gradient} \right), \qquad (19)$$
где $\gamma_{gradient} > 0$ – задаваемый коэффициент усиления.

Однако, как будет показано далее в рамках проведения компьютерного моделирования, точность оценивания параметров в случае применения нового подхода превосходит точность классического градиентного вида (19).

## Моделирование

Для иллюстрации работоспособности нового подхода проведено численное моделирование в пакете прикладных программ Matlab. Новый метод оценивания неизвестного параметра регрессионной модели (1) сравнивается с градиентным методом (19). Параметры при моделировании выбраны следующим образом: $\theta = 2$ – оцениваемый параметр, $\hat{\theta}(0) = 1.8$ – начальные условия оценки параметра, $\phi(t) = \sin t$ – регрессор, $k = 1$ – коэффициент фильтра, $\beta_1 = 3$, $\beta_2 = 5$ – коэффициенты фильтров в методе ДРСР, $\gamma_{gradient}$ – коэффициент усиления в методе градиентного спуска (19), $\kappa$ – коэффициент усиления в (18). В процессе моделирования сравниваются получаемые оценки параметра регрессионной модели (3) предлагаемым методом (на рисунках обозначены $\hat{\theta}_{new}$) и методом градиентного спуска (19) (на рисунках обозначены $\hat{\theta}_{gradient}$). Исследуется работа алгоритмов оценивания при различных видах помех измерения.

На Рис. 1 представлена помеха измерения $\overline{\delta(t)} = \sin 10t$ и сигнал $\delta(t)$.

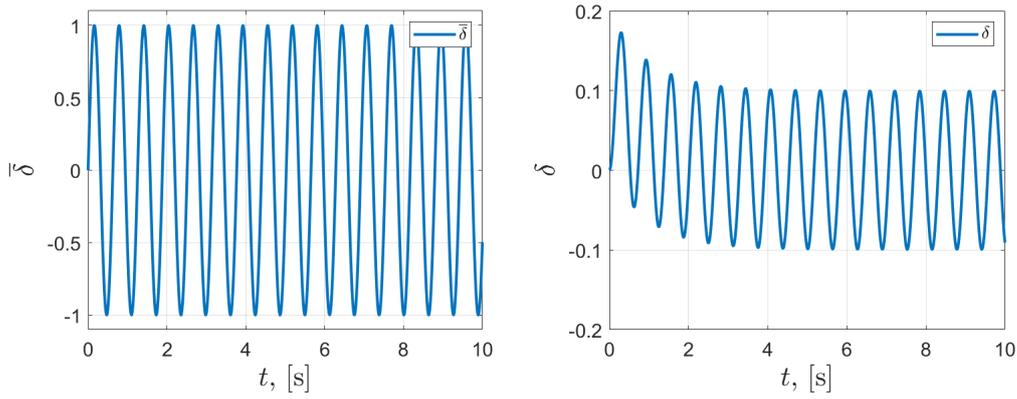

Рис. 1 — Помеха измерения $\overline{\delta(t)} = \sin 10t$ и сигнал $\delta(t)$

На Рис. 2 представлена работа алгоритмов при помехе измерения $\overline{\delta(t)} = \sin 10t$ и коэффициентах $\gamma_{gradient} = 10000$ и $\kappa = 10^8$.

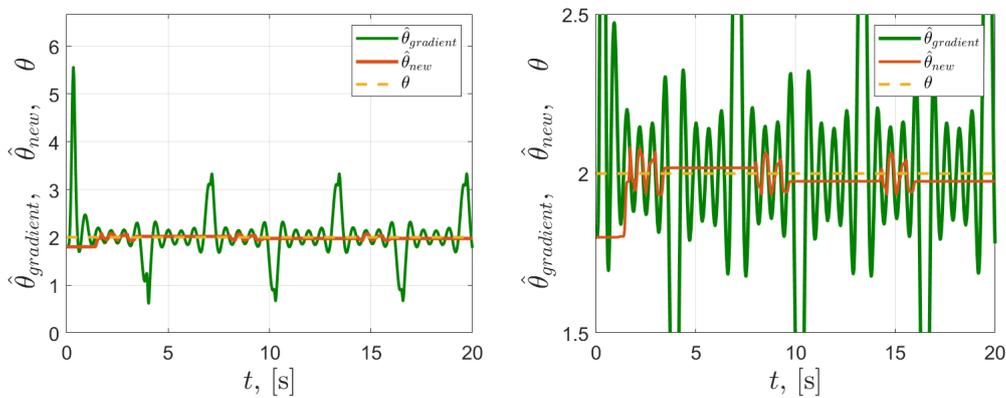

Рис. 2 — Графики оценки параметра $\theta$ предлагаемым в работе методом и градиентным методом при наличии в регрессии помехи оценивания $\overline{\delta(t)} = \sin 10t$

На Рис. 3 представлена помеха измерения $\overline{\delta(t)} = 0,5$ и сигнал $\delta(t)$.

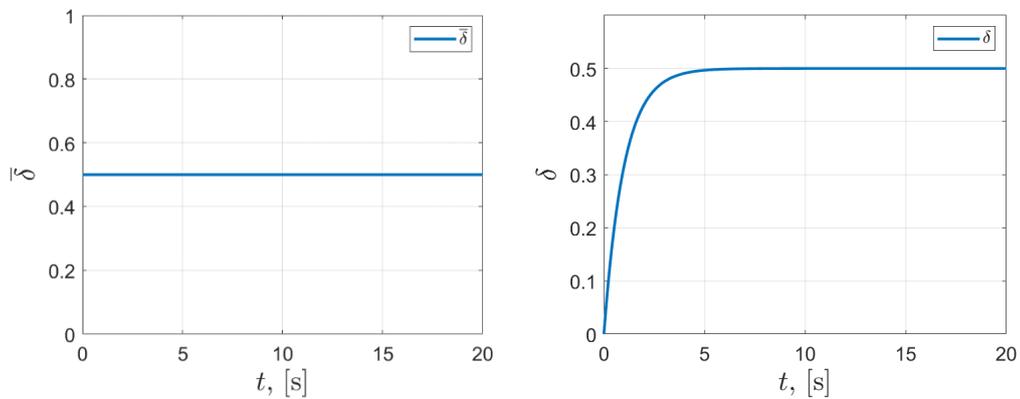

Рис. 3 — Помеха измерения $\overline{\delta(t)} = 0,5$ и сигнал $\delta(t)$

На Рис. 4 представлена работа алгоритмов при помехе измерения $\overline{\delta(t)} = 0{,}5$ и коэффициентах $\gamma_{gradient} = 10000$ и $\kappa = 10^8$.

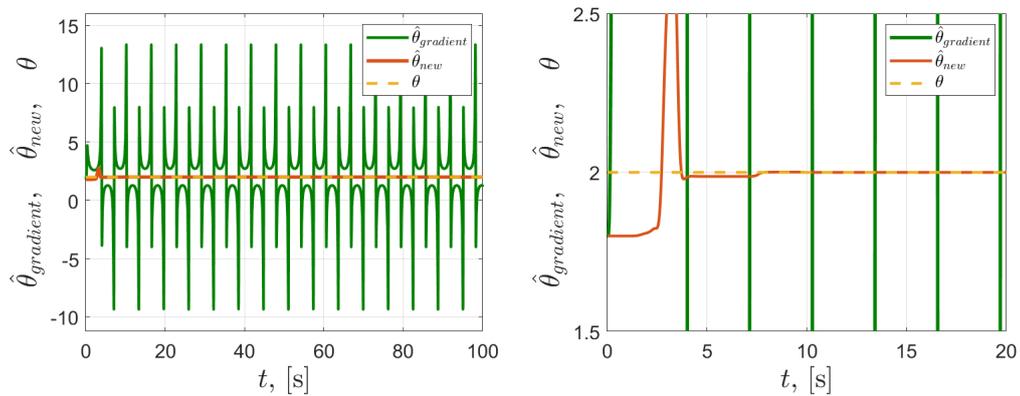

Рис. 4 — Графики оценки параметра $\theta$ предлагаемым в работе методом и градиентным методом при наличии в регрессии помехи оценивания $\overline{\delta(t)} = 0{,}5$

На Рис. 5 представлена помеха измерения $\overline{\delta(t)}$, заданная в виде равномерного случайного распределения на интервале $(-0.5; 0.5)$, и соответствующий ей сигнал $\delta(t)$.

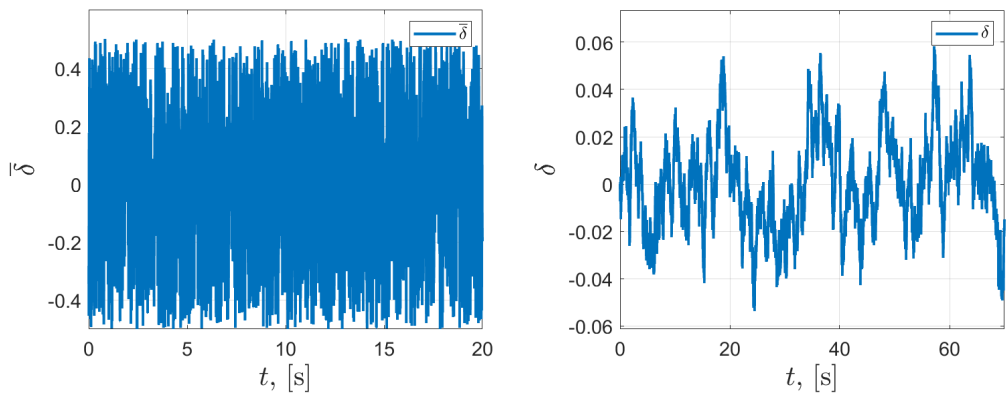

Рис. 5 — Помеха измерения $\overline{\delta(t)}$, заданная в виде равномерного случайного распределения на интервале $(-0.5; 0.5)$, и сигнал $\delta(t)$

На Рис. 6 представлена работа алгоритмов при помехе измерения $\overline{\delta(t)}$, заданной в виде равномерного случайного распределения на интервале $(-0.5; 0.5)$ и коэффициентах $\gamma_{gradient} = 10$ и $\kappa = 10^{10}$.

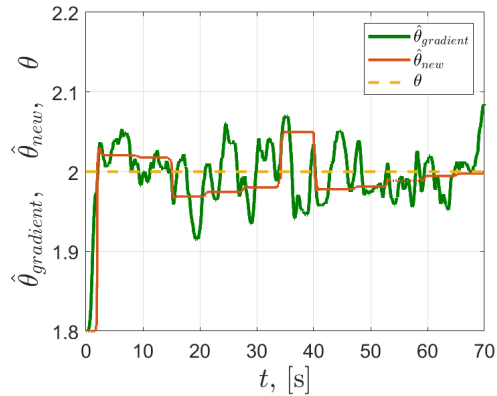

Рис. 6 — Графики оценки параметра $\theta$ предлагаемым в работе методом и градиентным методом при наличии в регрессии помехи оценивания $\overline{\delta(t)}$, заданной в виде равномерного случайного распределения на интервале $(-0.5; 0.5)$

Как следует из графиков переходных процессов, для трех исследованных случаев значения оценки $\hat{\theta}_{new}$ параметра $\theta$, полученные предлагаемым в работе методом, превосходят по точности (а в некоторых случаях существенно лучше) оценки $\hat{\theta}_{gradient}$, полученные методом градиентного спуска (19).

## Заключение

В статье представлен новый метод оценивания параметра скалярной регрессионной модели вида (1), содержащей помехи измерений иливозмущение. Задача идентификации параметра была решена с помощью получения новой регрессионной модели посредством применения нелинейного оператора, представляющего собой экспоненциальную функцию. Использование подобного преобразования, как было показано в статье, позволило существенно нивелировать влияние помехи измерения на оценку неизвестного параметра.

Для иллюстрации работоспособности предложенного подхода было проведено компьютерное моделирование с использованием пакета Matlab. В ходе моделирования было показано преимущество в точности оценивания параметра регрессионной модели предложенным методом по сравнению с методом градиентного спуска.

## Литература